\documentclass[conference]{IEEEtran}

\usepackage{cite}
\usepackage{array}
\usepackage{url}
\usepackage{graphicx}
\usepackage{tabularx} 
\usepackage{multirow}
\usepackage{color}
\usepackage{numprint} 
\npthousandsep{,}
\usepackage{arydshln} 
\usepackage{balance} 
\usepackage{todonotes}
\usepackage{blindtext}
\usepackage[english=american]{csquotes} 
\usepackage{listings}

\usepackage{framed}

\usepackage{numprint}
\npthousandsep{,}

\begin{document}

\title{Are There Functionally Similar Code Clones \\ in Practice?}

\author{\IEEEauthorblockN{Verena K\"afer}
\IEEEauthorblockA{University of Stuttgart, Germany\\
verena.kaefer@informatik.uni-stuttgart.de\\
https://orcid.org/0000-0002-7070-4519}
\and
\IEEEauthorblockN{Stefan Wagner}
\IEEEauthorblockA{University of Stuttgart, Germany\\
stefan.wagner@informatik.uni-stuttgart.de\\
https://orcid.org/0000-0002-5256-8429}
\and
\IEEEauthorblockN{Rainer Koschke}
\IEEEauthorblockA{University of Bremen, Germany\\
koschke@uni-bremen.de\\
https://orcid.org/0000-0003-4094-3444}
}

\maketitle

\begin{abstract}
Having similar code fragments, also called \emph{clones}, in software systems can lead to unnecessary comprehension, 
review and change efforts. Syntactically similar clones can often be encountered in practice. The same is not clear for 
only \emph{functionally} similar clones (FSC).

We conducted an exploratory survey among developers to investigate whether they encounter functionally similar
clones in practice and whether there is a difference in their inclination to remove them to syntactically similar clones.

Of the 34 developers answering the survey, 31 have experienced FSC in their professional work, and 24 have experienced
problems caused by FSCs. We found no difference in the inclination and reasoning for removing FSCs and syntactically
similar clones.

FSCs exist in practice and should be investigated to bring clone detectors to the same quality as for syntactically similar
clones, because being able to detect them allows developers to manage and potentially remove them.
\end{abstract}


\section{Introduction}
\label{sec:introduction}

Having similar code fragments, also called \emph{clones}, in software systems can lead to unnecessary comprehension, 
review and change efforts, and are a major source of faults~\cite{2009_juergens_inconsistens_clones}. A \emph{code clone} are at least two pieces of code that are similar according to a definition of similarity.
Most commonly, clone detection approaches look for exact clones, clones
with simple changes such as renaming, or even clones with additional changes. They can be detected by several detection approaches and tools~\cite{juergens2010code}.
 Yet, Juergens, Deissenboeck and Hummel~\cite{juergens2010code} were able to show  that these approaches are restricted
to detecting copy\&paste code. 
To address this, several researchers have proposed to abstract from the concrete syntax to detect what they call
\emph{semantic clones}.

The problem with classic approaches based on text, tokens, or syntax is that they cannot find clones with a completely different structure. We include those in the so called \emph{Functionally Similar Clones (\emph{FSCs})}~\cite{wagner:peerj16}. FSCs have the same or similar functionality but were generally created independently. Alternatively, they are sometimes the result of copy\&paste and additionally applied transformations rendering them syntactically dissimilar. They are also called \emph{type-4 clones}. Yet, type-4 clones are sometimes defined as having equivalent functionality. To make explicit that we also allow functional similarity and not only identity, we use the term FSC.

Whether two programs behave equivalently is undecidable in general. There may be cases, however, in which criteria exist that can be decided by an algorithm. But even those criteria are difficult to compute and therefore, only few approaches exist to detect FSCs. Additionally, there is little knowledge about whether these functionally similar clones exist in real projects, whether developers are aware of them, and whether they cause problems. 


We ran an open survey asking developers about their experiences with FSCs, how they handle them and how they would handle some given examples.


In contrast to related work we did not try to actually find FSCs but to see how they appear in practice and what developers think about them. 

We received 34 complete or almost complete responses to our survey.
Our results show that FSCs indeed exist in practice and give insights into the reasons why they exist and why they might or might not be refactored. 

We will first give an introduction into related work. Then we will describe our study design and present our results. The paper ends with a description of our threats to validity and a conclusion.

\section{Related Work}
\label{sec:related-work}

Krinke~\cite{Krinke:2001ev} proposed to use program dependence graphs (PDG) for finding semantic clones. 
Komondoor and Horwitz~\cite{Komondoor:2001it} also use PDGs for clone
detection and see the possibility to find non-contiguous clones as a main benefit. 
Gabel, Jiang and Su~\cite{Gabel:2008je} then combined abstract syntax trees with the analysis of dependence graphs in a tool called \textit{DECKARD} to better scale the approach. 
Krutz and Shihab~\cite{Krutz:2013ko} build on the control flow graph to create \enquote{concolic output} which they compare for similarity in their tool \textit{CCCD}. 
Marcus and Maletic~\cite{Marcus:2001dk} use the information retrieval technique \emph{latent semantic indexing} to
statically detect semantic similarities between code segments.
A very different approach to detect semantic clones comes from Kim et al.
\cite{Kim:2011jh} who use static analysis to extract the memory states for each procedure exit point. 
  
Jiang and Su~\cite{Jiang:2009bv} have been the first to
detect functionally similar code by using random tests and comparing the output. Hence, they were also the first to be able to detect
clones without any syntactic similarity. They were able to detect a
high number of functionally equivalent clones in a sorting benchmark
and the Linux kernel. Several of the detected clones are dubious,
however, as they only overlap in little functionality and Jiang and Su
doubt their usefulness themselves.

Deissenboeck et al.\ extended this approach to object-oriented systems in Java with more flexible chunking~\cite{Deissenboeck:2012gy}.
In contrast to Jiang and Su, they were able to identify only few FSCs in Java open-source systems.
Su et al.~\cite{su2016} improved this approach again and found several hundred FSCs in over a hundred open-source systems. Suzuki et al. then also worked on finding functional redundancy in code repositories~\cite{Suzuki:2017}. 

Yet, it is still not clear whether practitioners have experienced FSCs and whether they would consider them as interesting to
refactor as other clones.

Finally, Wagner et al. found that less than 16\,\% of FSC pairs have actual syntactic similarities~\cite{wagner:peerj16}. They provide a
benchmark for FSCs which has, to the best of our knowledge, not yet been used to test FSC detection approaches.

\section{Study Design}
\label{sec:study-design}

To better understand FSCs in practice, we decided to conduct a survey
so that we can reach a broad sample of developers covering different companies and domains. As we have almost
no clear theory on FSCs in practice, we designed an \emph{exploratory} survey. We describe
the research questions that guided the design, a hypotheses that we test on the data, the data collection procedure
including the questionnaire, the analysis procedure and the validity procedure in the following.

\subsection{Research Questions}

Our overall research goal is to better understand FSCs in practice. This includes whether this phenomenon
occurs in practice, whether practitioners are aware of it, and how they handle FSCs. We are especially interested
in whether FSCs are handled differently than other (type 1--3) clones. We break this down into the following
research questions:

\noindent\textbf{RQ 1: What experience do practitioners have with FSCs?}

\noindent\textbf{RQ 2: How are FSCs handled in practice?}

\noindent\textbf{RQ 3: Are FSCs handled differently than other clones?}

\subsection{Hypothesis}
\label{sec:hypotheses}

Although this is an exploratory study and we mostly analyze the data with descriptive statistics and qualitative
methods, we formulate one hypothesis which we had as part of the motivation to do the study. This hypothesis
is about differences in whether developers would refactor a clone depending on the syntactic similarity of the
clone instances. In other words: Is there a difference in refactoring
behavior w.r.t.\ FSCs and type-1--3 clones?

\noindent$H_0$: \emph{There is no difference in the inclination of developers to refactor a clone between syntactically 
similar and syntactically dissimilar clones.}

\noindent$H_A$: \emph{There is a difference in the inclination of developers to refactor a clone between syntactically 
similar and syntactically dissimilar clones.}

\subsection{Data Collection Procedures}
\label{sec:data-collection-procedures}

We provide the full questionnaire openly on Zenodo~\cite{clone-survey-data}. It contains 17 closed and 9 open questions.
The open questions reflect the exploratory nature of the survey so that we can better understand the background of the
answers to the closed questions. In a first block, we ask the participants about their prior knowledge 
regarding FSCs. This block aims to answer part of RQ 1. A second block contains example code of four clones. The examples were created by one of the authors. 
The one in Listing~\ref{lst:sort} shows two sorting methods. Listing~\ref{lst:reverse} shows two methods to reverse strings. In Listing~\ref{lst:strings} we can see two classes with string manipulation methods.  The last clone in Listing~\ref{lst:distance} shows two methods calculating the distance between two points, one in 2D and one in 3D.

We chose them deliberately in a way that two clones (\textit{Sort} in Listing~\ref{lst:sort} and \textit{Reverse} in Listing~\ref{lst:reverse}) are
syntactically dissimilar---and hence probably developed independently---while the clones \textit{StringHandling} (Listing~\ref{lst:strings})
and \textit{Distance} (Listing~\ref{lst:distance}) are more syntactically similar. Hence, they might have been created by copy\&paste and
could probably be found by a type-3 clone detector.

We ask for each
presented clone pair whether the participants would refactor this clone and why. Furthermore, we ask them if they would
like to see this clone in a tool for specific reasons (e.g.,\ to improve maintainability or prevent bugs). This block shall answer
the RQs 2 and 3. In the third block, we ask for more detailed experiences with FSCs. For example, we ask whether FSCs
are problems in the participants' projects. This block helps to answer RQ 1 and RQ 2. The last block of questions contains
demographic questions.

\definecolor{javared}{rgb}{0.6,0,0} 
\definecolor{javagreen}{rgb}{0.25,0.5,0.35} 
\definecolor{javapurple}{rgb}{0.5,0,0.35} 
\definecolor{javadocblue}{rgb}{0.25,0.35,0.75} 

\lstset{language=Java,
	basicstyle=\fontsize{8}{8}\selectfont,
	keywordstyle=\color{javapurple}\bfseries,
	stringstyle=\color{javared},
	commentstyle=\color{javagreen},
	morecomment=[s][\color{javadocblue}]{/**}{*/},
	tabsize=4,
	showspaces=false,
	showstringspaces=false}

 \begin{lstlisting}[caption={Clone example: sorting arrays (syntactically dissimilar)}, label={lst:sort}]
 /**
 * @param intArr
 * @return sorts the input array using bubble sort
 */
 public int[] bubbleSort(int[] intArr) {
    int k;
    for (int i = 0; i < intArr.length - 1; i++) {
       if (intArr[i] < intArr[i + 1]) {
          continue;
       }
       k = intArr[i];
       intArr[i] = intArr[i + 1];
       intArr[i + 1] = k;
       bubbleSort(intArr);
    }
    return intArr;
 }
 
 /**
 * @param intArr
 * @return sorts the given array using insertion sort
 */
 public int[] insertSort(int[] intArr) {
    int k;
    for (int i = 0; i < intArr.length; i++) {
       for (int j = intArr.length - 1; j > 0; j--) {
          if (intArr[j - 1] > intArr[j]) {
             k = intArr[j];
             intArr[j] = intArr[j - 1];
             intArr[j - 1] = k;
          }
       }
    }
    return intArr;
 }
 \end{lstlisting}

\begin{lstlisting}[caption={Clone example: reversing strings (syntactically dissimilar)}, label={lst:reverse}]
/**
* @param input
* @return the reversed input string
*/
public String reverse(String input) {

   StringBuffer buffer = 
      new StringBuffer(input.length());
   for (int i = input.length() - 1; i >= 0; i--) {
      buffer.append(input.charAt(i));
   }
   return buffer.toString();
}

/**
* @param str
* @return the reversed input string or null, if the 
* input is null
*/
public static String reverse1(final String str) {
   if (str == null) {
      return null;
   }
   return new StringBuilder(str).reverse().toString();
}
\end{lstlisting}

\begin{lstlisting}[caption={Clone example: manipulating strings (syntactically similar)}, label={lst:strings}]

public static class StringHandling {

   /**
   * @param input
   * @return the first letter of the input string
   */
   public static String getFirstLetter(String input) {
      return String.valueOf(input.charAt(0));
   }

   /**
   * @param s1
   * @param s2
   * @return the sum of the strings' integer values
   */
   public static int sumOfStrings(String s1, 
         String s2) {
      return Integer.valueOf(s1) + Integer.
         valueOf(s2);
   }

}

public class StringHandler {

   /**
   * @param input
   * @return the first letter of the input as String
   */
   public String getFirstCharacter(String input) {
      return Character.toString(input.charAt(0));
   }

   /**
   * @param s1
   * @param s2
   * @return the sum of the input strings
   */
   public int sum(String s1, String s2) {
      return Integer.parseInt(s1) + Integer.
         parseInt(s2);
   }

   /**
   * @param s
   * @return the value of the input multiplied by -1
   */
   public int negativeValueOfString(String s) {
      return Integer.parseInt(s) * -1;
   }
}
\end{lstlisting}

\begin{lstlisting}[caption={Clone example: calculating the distance between two points (syntactically similar)}, label={lst:distance}]

public static double calculateDistance2D(Point p1, 
      Point p2) {
   int xDiff = p2.x - p1.x;
   int yDiff = p2.y - p1.y;
   int sum = xDiff ^ 2 + yDiff ^ 2;
   return Math.sqrt(sum);
}

public static double calculateDistance3D(Point p1, 
      Point p2) {
   int xDiff = p2.x - p1.x;
   int yDiff = p2.y - p1.y;
   int zDiff = p2.z - p1.z;
   int sum = xDiff ^ 2 + yDiff ^ 2 + zDiff ^ 2;
   return Math.sqrt(sum);
}
\end{lstlisting}

We used Rogator to implement the questionnaire. One of the authors implemented a first version of the questionnaire and then improved it in several cycles based on the suggestions of the other authors and some test runs with colleagues. The questionnaire was then openly distributed via email to previous industry contacts, Twitter and at an industrial conference for practitioners. It was online for about six months.

\subsection{Analysis Procedures}

We have analyzed the participants' answers using R. We use bar charts to show the frequencies and distributions of the answers. 
To be on the safe side, we treat the answers to the refactoring questions as ordinal data. Hence, we use the Wilcoxon sum rank
test to test our hypothesis. We also calculate Cohen's d for the effect size. We use all the data from both syntactically dissimilar FSCs as one population
and all the data from both syntactically similar clones as a second population and compare them. We use a confidence level of 5\,\%.

We provide the complete R analysis script on Zenodo~\cite{clone-survey-data}.

We selected representative answers to the qualitative questions and present them as citations related to the quantitative bar charts.

\subsection{Validity Procedures}
\begin{figure*}[htb]
	\begin{center}
		\includegraphics[width=.9\textwidth]{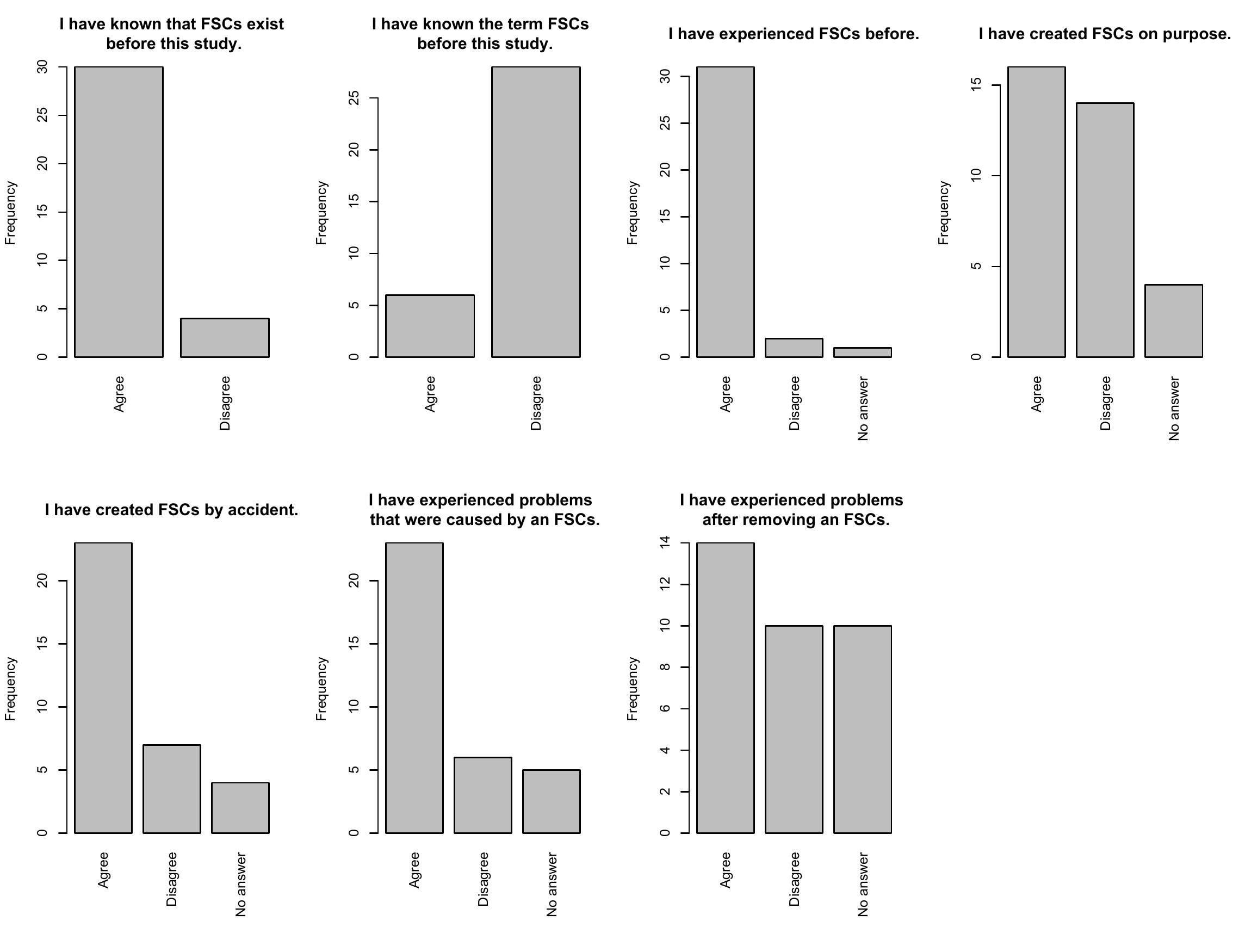}
		\caption{Results for general statements on FSCs}
		\label{fig:general-statements}
	\end{center}
\end{figure*}

We restrict our analysis to existential questions (``Do FSCs occur in practice?'') and relationships between the
answers of the respondents (``Is there a difference in handling FSCs and other clones?''). We explicitly refrain from 
generalizing the distributions of the answers to a large population as we have a relatively small sample size and no
clear sampling frame for a larger population such as all software developers. This preliminary study is intended to explore whether there is initial evidence that FSCs are worthwhile to be investigated in a series of more detailed and comprehensive studies.

\section{Study Results}
\label{sec:study-results}

Overall, we had 122 responses to our survey. Unfortunately, we had to discard 88 responses, because they answered
only a few questions. This leaves us with 34 fully or almost fully answered questionnaires. Of our respondents, 33 are male,
only 1 female. Twenty-three of the respondents live in Germany, one in Russia, one in Austria and one in the UK. Most
of the respondents have more than 10 years of experience with programming. Four have 6 to 10 years, one 2 to 5 years
and one less than a year of experience. The respondents considered a wide variety of programming languages, their
\emph{main} programming language including Java, C, C++, C\#, Python, Ruby, Go, Scala, Ada, TypeScript, Haskell,
PHP, Rust and Swift.

We provide the full dataset of the survey responses as open data on Zenodo~\cite{clone-survey-data}. We will
structure the results according to the research questions, provide the quantitative results first and use the qualitative
results to further describe and interpret the results.

\subsection{RQ 1: Prior Knowledge of FSCs}

Figure~\ref{fig:general-statements} shows aspects on how FSCs were perceived before our study. Interestingly, 30 participants have known that FSCs exist before answering the questionnaire, 31 have experienced them. Twenty-eight participants have not known the term \textit{Functionally Similar Clone} before, which leads to the conclusion that the phenomenon of functionally similar code fragments exists, but is not commonly referred to with the term \textit{Functionally Similar Clone}. Hence, we observe from the data that (a) FSCs exist in projects in practice and (b) developers are aware of them. They might not have a name for them, but they know that there are functionally similar code fragments. 

Our results also show that FSCs can easily be created, either on purpose (16 participants) or by accident (23 participants). As reasons for creating FSCs on purpose the participants stated that they \emph{\enquote{had to add FSCs because of the compatibility and the performance}}, because the used programming language requires different implementations for comparing elements or to \emph{\enquote{exchange [the] proven working solution by [a] better version and test the new one against the old one}}, amongst others. 

Unintentional FSCs were created due to \emph{\enquote{developers not talking to each other}}, because they \emph{\enquote{didn't know the other existed}} or because there were \emph{\enquote{no common infrastructure components}}.
Purposely created FSCs might have benefits, like a better performance, and the developers know that they exist. More problematic are unknown FSCs, because errors might be done twice and often there might be a better utility method which solves the problem in an efficient way but is not known to the developer. 

\newpage
In either way, 23 participants have experienced problems that were caused by an FSC before. Unfortunately, just removing the clone is no safe solution, as 14 participants have experienced problems after removing an FSC, whereas 10 participants did not. 10 participants have not answered this question, which could mean that some participants have never removed an FSC. When maintaining code there is always the chance to break something. In the case of FSCs, it might be dangerous to just remove a code fragment if the fragments were not functionally identical, for instance, when one fragment included some exception handling that did not exist in its counterpart.

\subsection{RQ 2: Handling of FSCs in Practice}

\begin{figure*}[htb]
\begin{center}
\includegraphics[width=.8\textwidth]{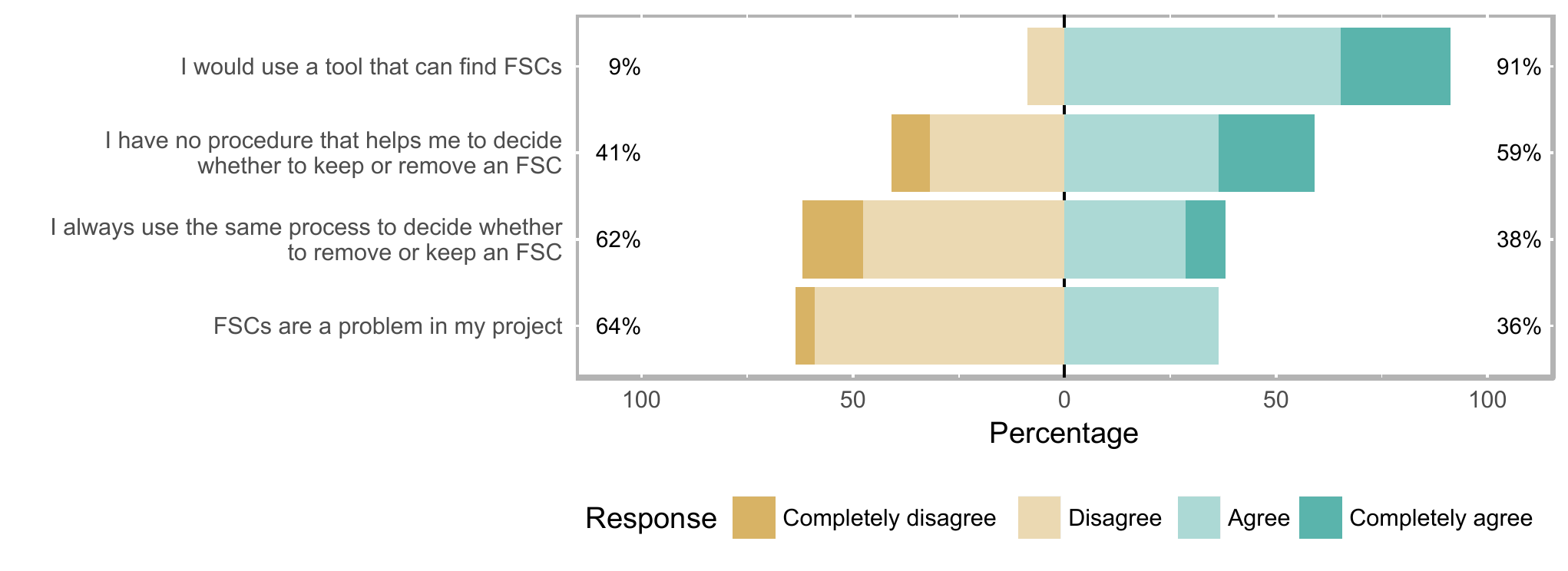}
\caption{Experiences and expectations with FSCs in practice}
\label{fig:experience}
\end{center}
\end{figure*}

In practice, there are many ways to handle FSCs. Figure~\ref{fig:experience} shows that 91\,\% of the participants would use a tool that can find FSCs. In their current situation, 59\,\% of the participants have no procedure to decide whether to keep or remove an FSC, 41\,\% do have such a procedure. It looks like many developers individually decide whether to remove or to keep an FSC without standard criteria. Interestingly, 41\,\% do have a procedure that helps them to decide how to handle an FSC. It might be interesting to have a look at these processes and to see what might help other developers decide. 

Anyhow, 62\,\% of the participants do not always use the same process to decide whether to keep or remove an FSC, whereas 38\,\% do have such a process. 
They seem to decide individually, depending on the situation and do not always do it in the same way. 

Luckily, only about one third (36\,\%) of the participants has experienced a problem in their current project regarding FSCs. The number is lower than the number of developers that have experienced problems caused by FSCs before, which could mean that such problems do not occur in every project and the former experiences could have happened in an earlier project. As reasons to keep a known FSC the participants stated \emph{\enquote{distributed responsibility}}, \emph{\enquote{bad test coverage [or the] risk of unexpected side effects in code that depends on the FSC}}, amongst others. Additionally, they stated the reasons that were also presented for creating FSCs on purpose. 
For removing existing FSCs, the participants argued with \emph{\enquote{better maintainability}} or wanting to \emph{\enquote{reduce the complexity and dangerous duplication}}. As we can see, there are valid reasons for code fragments with the same functionality, as long as they have a defined purpose. If this purpose is not given, many participants tend to remove the clone to improve maintenance and cure other bad smells.

\subsection{RQ 3: Difference Between FSCs and Other Clones}

\subsubsection{Descriptive Statistics}

\begin{figure}[htb]
	\begin{center}
		\includegraphics[width=.45\textwidth]{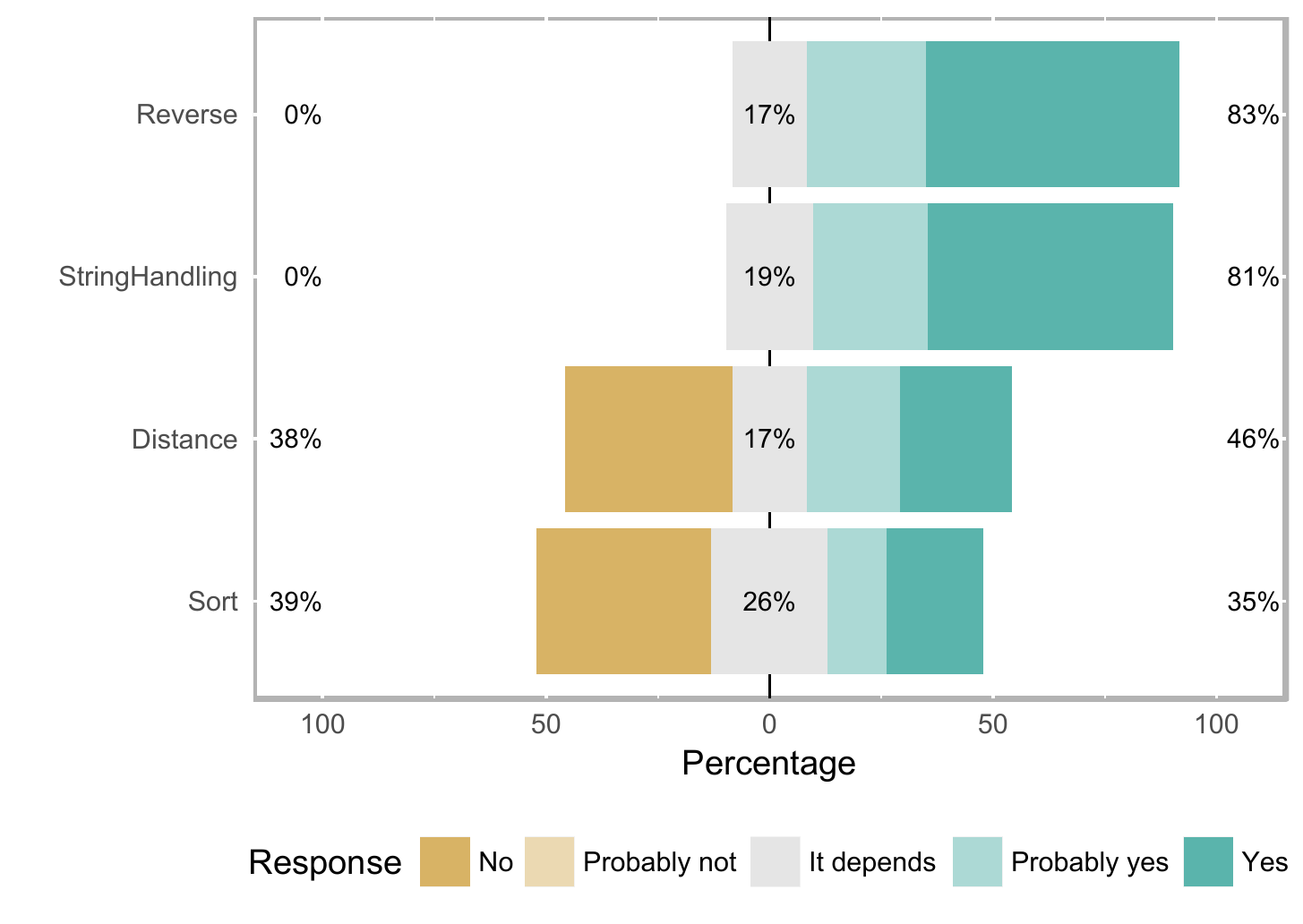}
		\caption{I would refactor this code fragment.}
		\label{fig:refactor}
	\end{center}
\end{figure}

When looking at the four clone examples, the participants first had to decide if they would refactor the clones (Figure~\ref{fig:refactor}). For the string reversing clone example and the string handling classes about 80\,\% of the participants agreed to get rid of the clones. The other participants decided based on the situation. None of the participants wanted to definitely not refactor these clones. 
For the distance calculating methods and the sort algorithms the answers were more diverse. About 40\,\% wanted to keep both fragments, the rest either wanted to refactor the code or decide depending on the situation. 
We think that it was easier for the participants to see whether the string handling clones provided similar functionality and therefore decided to get rid of them than for the other two clones. For the distance calculation it was not clear, whether the functionality is always the same as one distance was calculated in 2D and one in 3D. The same is true for the sorting algorithms. As many participants referred to performance in other questions, we think that it might be hard to decide whether the sorting algorithms really provide the same performance for different inputs. 
Another reason might be the difficulty to clearly check whether both implementations fully implement the needed functionality.

\begin{figure}[htb]
\begin{center}
\includegraphics[width=.45\textwidth]{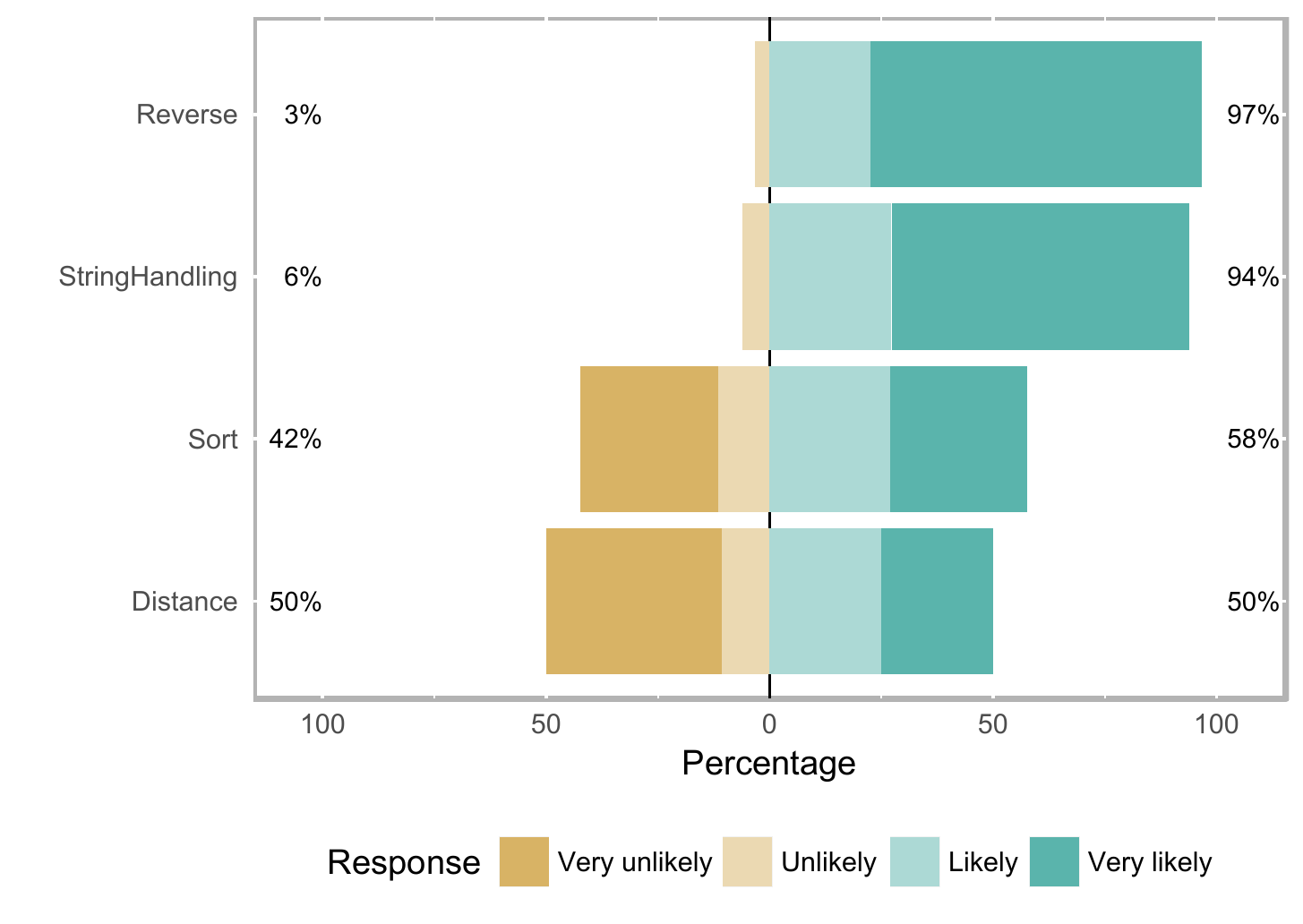}
\caption{I would like to see this clone in a tool because I want to get rid of one of the code fragments.}
\label{fig:get-rid}
\end{center}
\end{figure}

In Figure~\ref{fig:get-rid} we can see whether the participants would like to find a clone to get rid of one of the fragments. The results are similar to the previous question regarding refactoring the clone. For the \textit{StringHandling} clones, about 95\,\% would get rid of one fragment, for the other two about one half would keep both fragments and the other half would get rid of one.

\begin{figure}[htb]
\begin{center}
\includegraphics[width=.45\textwidth]{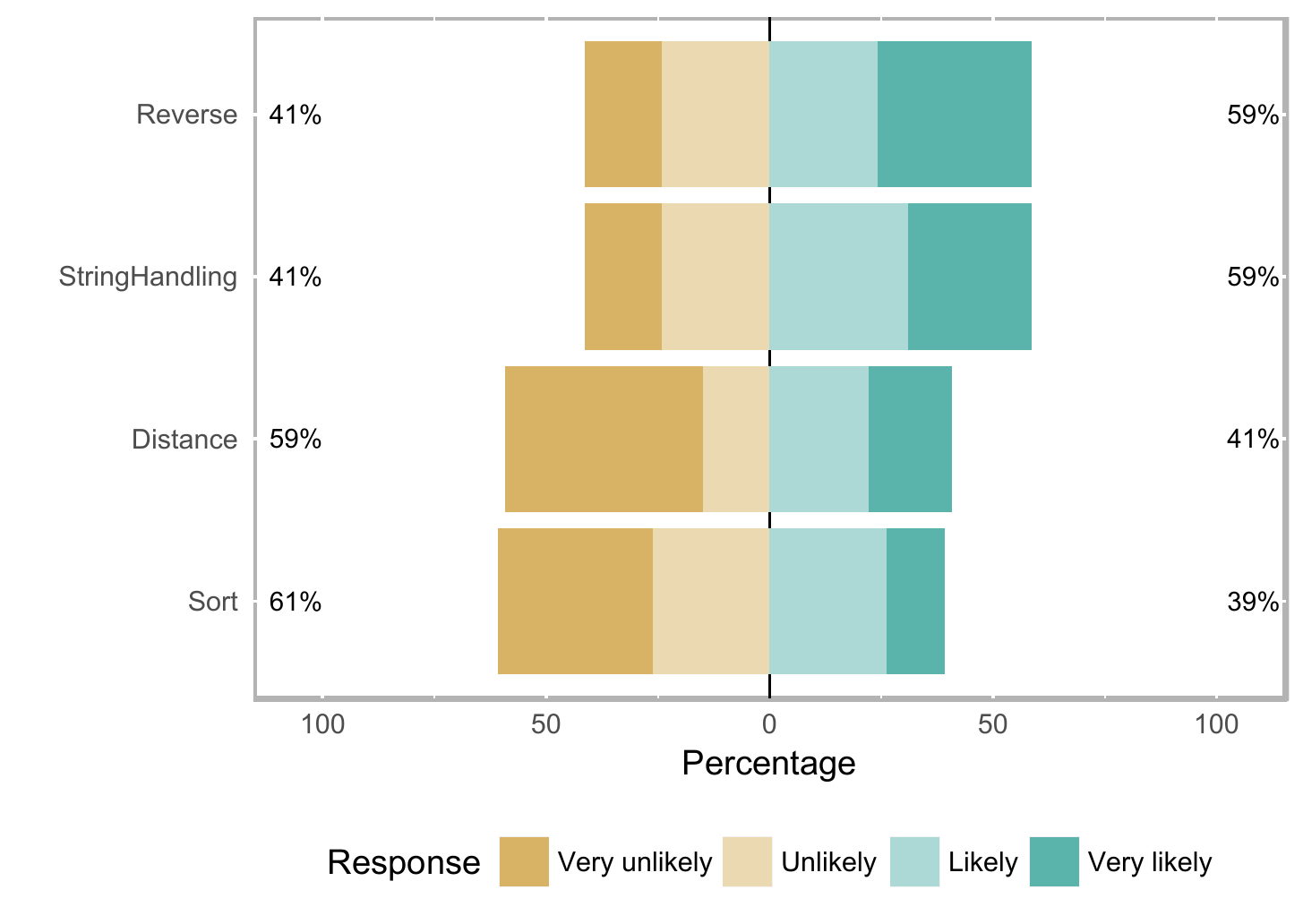}
\caption{I would like to see this clone in a tool because I want to trace changes.}
\label{fig:trace}
\end{center}
\end{figure}

For the goal of tracing changes, the results are more balanced (Figure~\ref{fig:trace}). About 60\,\% would like to see the \textit{StringHandling} clones in a tool to trace changes. Apparently tracing changes is not as interesting for these clones as getting rid of one. Also for the other two clones fewer participants would use a tool for this purpose.

\begin{figure}[htb]
\begin{center}
\includegraphics[width=.45\textwidth]{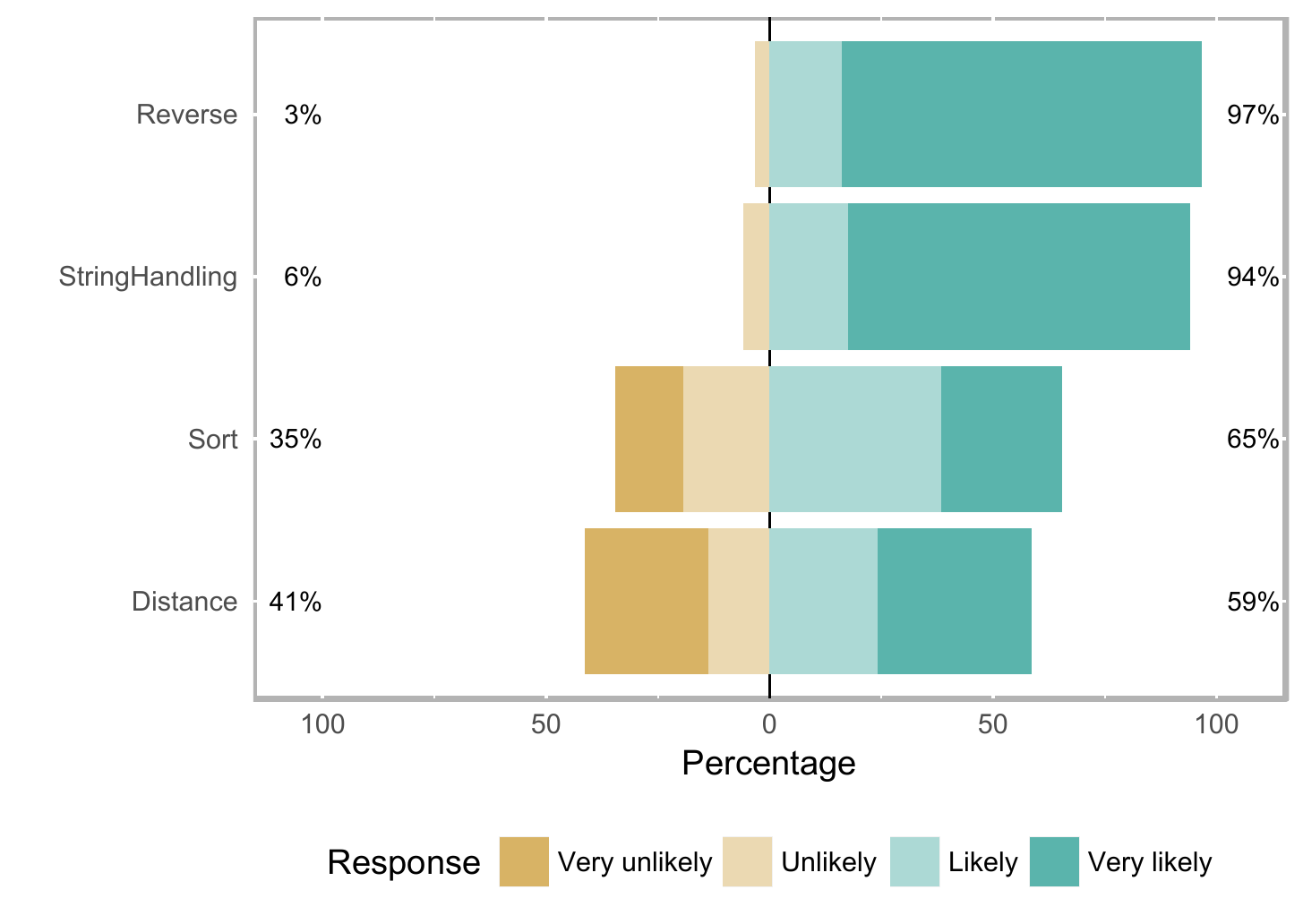}
\caption{I would like to see this clone in a tool because I want to improve the maintainability.}
\label{fig:maintain}
\end{center}
\end{figure}

In contrast to this, many participants would like to see the provided clones in a tool to improve the maintainability (Figure~\ref{fig:maintain}). It looks like getting rid of a clone is as important as improving the maintainability, because for the \textit{StringHandling} clones the percentages are the same. For the other two clones, the wish of a better maintainability is even higher than the wish of getting rid of one clone fragment.

\begin{figure}[htb]
\begin{center}
\includegraphics[width=.45\textwidth]{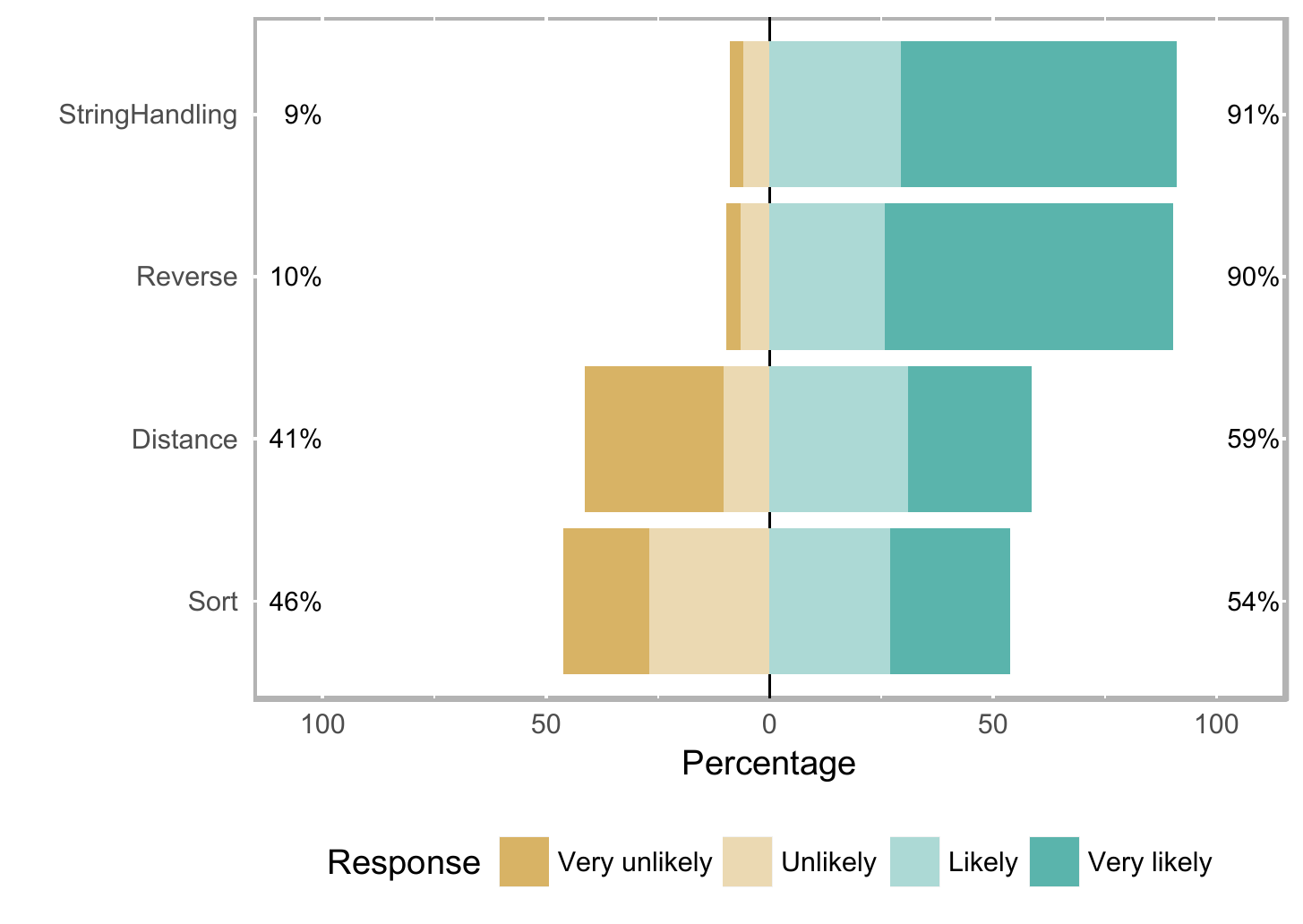}
\caption{I would like to see this clone in a tool because I want to prevent bugs.}
\label{fig:bugs}
\end{center}
\end{figure}

Preventing bugs is nearly as popular as improving the maintainability (Figure~\ref{fig:bugs}). It might again be seen as a part of maintainability. Slightly fewer participants would use a tool for this, but still more than half of them would do it for the distance and sort clones and over 90\,\% for the \textit{StringHandling} clones. 

In general it is interesting to see that the two \textit{StringHandling} clones that most participants would refactor also had very similar values for the other questions. Also the other two clones had similar results. This leads to the conclusion that some clones can easily be recognized as similar and also easily refactored whereas other clones are not that easy to recognize and have a higher potential that they have been created on purpose for various reasons. 

The most interesting observation for us overall is that the answers for handling the clones do not seem to depend on their syntactic similarity. \textit{Reverse} and \textit{Sort} on the one side
and \textit{StringHandling} and \textit{Distance} on the other are not grouped in that way in the answers. On the contrary, \textit{Reverse} and \textit{StringHandling} always have more ``likely'' ratings than
\textit{Sort} and \textit{Distance}.

\subsubsection{Hypothesis Testing}
The hypothesis test resulted in the p-value 0.95 which is far over our $\alpha$ level of 5\,\%. Hence we cannot reject the null hypothesis that there is no difference in the inclination of developers to refactor a clone between syntactically similar and syntactically dissimilar clones. Also the effect size of 0.02 is negligible. 

Hence, the formal test further supports our observation that the syntactic similarities are not the main factor in deciding how to handle a clone. FSCs that have been developed independently and have little syntactic similarity can be as interesting as any other clone depending on context and the goals of the developers.

\section{Threats to Validity}
\label{sec:threats}

Our results are not generalizable, as our sample size is quite small. Also the participants do not come from one homogeneous crowd of software developers but voluntarily decided to take part in the survey when somehow receiving the link to the questionnaire. It can also be seen in the large drop-out rate that probably only developers really interested in this topic answered most questions. This is not a problem, because we did not aim to generalize. Our goal was to see whether there are experiences in practice with FSCs. 

Our results also rely on the truthfulness of the participants' answers. Remembering that one has seen such a clone is quite easy---remembering what this clone was about is more difficult. We do not count this as a big threat, because (a) the questionnaire was not only about experiences but also about concrete code samples, and (b) we did not see any invalid or non-precise answers. Therefore, we think that our results are still valid. 

\section{Conclusions}
\label{sec:conclusion}

We conducted an open survey to understand whether functionally similar code clones exist in practice, whether 
developers are aware of them and whether they handle clones that are syntactically dissimilar differently to syntactically
similar FSCs. We received 34 (almost) complete answers that allow us an exploratory quantitative and
qualitative analysis.

FSCs do occur in practice and developers are aware of them. We cannot estimate the distribution in a larger population
because of our small sample size and because respondents are probably interested because they might have already
seen an FSC in practice. Syntactic similarity is not the differentiating factor whether an FSC is refactored or not.
Functional suitability can be sufficient to decide to refactor but is also not sufficient.

Our results suggest that FSCs are worthwhile to be investigated further. Complementing the first studies investigating prototype FSC detectors, we think that automatic detection of all FSCs would be beneficial for practitioners. We believe there should be more research in that area to bring it closer to the level of precision and recall achieved in the detection of type-1--3 clones.
Additionally, some research about the reasons why FSCs get created might be interesting. As we could see, not all FSCs get created by accident, therefore we could have a look at how to communicate and document intended FSCs.

\balance

\section*{Acknowledgements}
We are grateful to all participants of our survey and all people who further distributed
it by forwarding our mails or retweeting our tweets.

\bibliographystyle{IEEEtran}

\bibliography{IEEEabrv,clone-survey}

\end{document}